\documentclass[prl,amsmath,amssymb,twocolumn,superscriptaddress,showpacs]{revtex4-1}

\usepackage{amsmath,amssymb}
\usepackage[usenames]{color}
\usepackage{amssymb}
\usepackage{grffile}
\usepackage[pdftex]{graphicx}
\usepackage{amsmath, amstext, amssymb, amsfonts, amsxtra}
\usepackage{textcomp}
\usepackage{xspace}
\usepackage{bbm}

\newcommand{\be}{\begin{equation}}
\newcommand{\ee}{\end{equation}}

\newcommand{\xiamen}{Department of Physics, Key laboratory of Low Dimensional
Condensed Matter Physics (Department of Education of Fujian Province), and
Jiujiang Research Institute, Xiamen University, Xiamen 361005, Fujian, China}
\newcommand{\como}{Center for Nonlinear and Complex Systems, Dipartimento di
Scienza e Alta Tecnologia, Universit\`a degli Studi dell'Insubria, via
Valleggio 11, 22100 Como, Italy}
\newcommand{\infn}{Istituto Nazionale di Fisica Nucleare, Sezione di Milano,
via Celoria 16, 20133 Milano, Italy}
\newcommand{\brazil}{International Institute of Physics, Federal University
of Rio Grande do Norte, Campus Universit\'ario - Lagoa Nova, CP. 1613, Natal,
Rio Grande Do Norte 59078-970, Brazil}
\newcommand{\NEST}{NEST, Istituto Nanoscienze-CNR, I-56126 Pisa, Italy}

\begin{document}

\title{Thermodynamic bound on heat to power conversion}
\author{Rongxiang Luo}
\affiliation{\xiamen}
\author{Giuliano Benenti}
\affiliation{\como}
\affiliation{\infn}
\affiliation{\NEST}
\author{Giulio Casati}
\affiliation{\como}
\affiliation{\brazil}
\author{Jiao Wang}
\affiliation{\xiamen}

\begin{abstract}
In systems described by the scattering theory, there is an upper bound, lower
than Carnot, on the efficiency of steady-state heat to work conversion at a
given output power. We show that interacting systems can overcome such bound
and saturate, in the thermodynamic limit, the much more favorable linear-response
bound. This result is rooted in the possibility for interacting systems to achieve
the Carnot efficiency at the thermodynamic limit without delta-energy filtering,
so that large efficiencies can be obtained without greatly reducing power.
\end{abstract}


\maketitle

{\it Introduction.}
The increasing energy demand and the depletion and environmental impact of fossil
fuels calls for renewable and eco-friendly energy resources. In this frame, nanoscale
thermal engines~\cite{GiazottoPekola2006, Shakouri2011, Dubi2011, Hanggi2011,
Sothmann2014, Benenti2016, MuhonenPekola2012, Seifert2012, Kosloff2013, Gelbwaser2015,
Vinjanampathy2015, Benenti2017} will play an important role and might become part of
the energetic mix of the future. A crucial point is the efficiency of such engines.
Given any heat engine operating between two reservoirs at temperature $T_L$ and
$T_R$ ($T_L>T_R$), the efficiency of energy conversion is upper bounded by the Carnot
efficiency $\eta_C=1-T_R/T_L$. This limit can be achieved for dissipationless heat
engines. Such ideal machines operate reversibly and infinitely slowly, and therefore
the extracted power vanishes in the Carnot limit. For any practical purpose it is
therefore crucial to consider the \emph{power-efficiency trade-off}, in order to
design devices that work at the maximum possible efficiency for a given output power.

For steady-state conversion of heat to work in quantum systems which can be modelled
by the Landauer-B\"uttiker scattering theory, this problem was solved theoretically
by Whitney~\cite{whitney1, whitney2}. Indeed, he found a bound on the efficiency at
a given output power $P$, which equals the Carnot efficiency at $P=0$, and decays
with increasing $P$. This upper bound is achieved when only particles within a given
energy window (determined by the desired output power $P$) of width $\delta(P)$ can
be transmitted through the system. The Carnot efficiency is obtained for delta-energy
filtering~\cite{mahan,linke1,linke2}, that is, when $\delta\to 0$, and in such limit
the output power vanishes. This interesting result establishes a bound for an important
class of systems. Now the relevant question is: how general is this bound? For general
interacting systems, can this bound be overcome, thus allowing for a better
power-efficiency trade-off?

In this letter, we give a positive answer to this question for classical systems.
Indeed, we show that interacting, nonintegrable momentum-conserving systems, overcome
the bound from the scattering theory. These systems can achieve the Carnot efficiency
at the thermodynamic limit, with a much more favorable power-efficiency trade-off than
allowed by the scattering theory. Therefore, interactions can significantly improve
the performance of steady-state heat to work conversion. This result is rooted in
the possibility, for interacting systems, to achieve the Carnot efficiency without
delta-energy filtering. Our results are illustrated by means of extensive numerical
simulations of classical models of elastically colliding particles.

{\it Classical reservoirs.}
For concreteness, we consider a one-dimensional system (even though, as discussed in
the conclusions and shown in Sec.~D of the supplementary material, our analysis can
be extended to higher dimensions), whose ends are in contact with left/right reservoirs,
characterized by temperature $T_\alpha$ and electrochemical potential $\mu_\alpha$
($\alpha=L,R$). The reservoirs are modelled as infinite one-dimensional ideal gases,
with particle velocities described by the Maxwell-Boltzmann distribution, $F_\alpha(v)=
\sqrt{\frac{m}{2\pi k_B T_\alpha}}\exp\left(-\frac{m v^2}{2 k_B T_\alpha}\right)$,
where $k_B$ is the Boltzmann constant and $m$ the mass of the particles. We use a
stochastic model of the reservoirs~\cite{carlos2001,carlos2003}: whenever a particle
of the system crosses the boundary which separates the system from the left or right
reservoir, it is removed. On the other hand, particles are injected into the system
from the boundaries, with rates $\gamma_\alpha$. The injection rate $\gamma_\alpha$
is computed by counting how many particles from reservoir $\alpha$ cross the
reservoir-system boundary per unit time:
$\gamma_\alpha=\rho_\alpha \int_0^\infty dv v F_\alpha(v)
= \rho_\alpha \sqrt{\frac{k_B T_\alpha}{2\pi m}}$,
with $\rho_\alpha$ the particle number density of the ideal gas in reservoir $\alpha$.
A standard derivation~\cite{saito2010} then shows that the density $\rho_\alpha$ is
related to the electrochemical potential $\mu_\alpha$ as follows:
$\mu_\alpha=k_B T_\alpha \ln(\rho_\alpha\lambda_\alpha)$,
where $\lambda_\alpha=h/\sqrt{2\pi m k_B T_\alpha}$ is the
de Broglie thermal wave length and $h$ the Planck constant.

{\it Non-interacting systems.}
In this case, the particle current reads~\cite{saito2010}
\begin{equation}
J_\rho=
\gamma_L\int_0^\infty d\epsilon u_L(\epsilon) {\cal T} (\epsilon)
-\gamma_R\int_0^\infty d\epsilon u_R(\epsilon) {\cal T} (\epsilon),
\label{eq:nonint1}
\end{equation}
where $u_\alpha(\epsilon)=\beta_\alpha e^{-\beta_\alpha\epsilon}$, with
$\beta_\alpha=(k_B T_\alpha)^{-1}$, is the energy distribution of the particles injected
from reservoir $\alpha$ and ${\cal T}(\epsilon)$ is the transmission probability for
a particle with energy $\epsilon$ to transit from one end to another of the system,
$0\le {\cal T}(\epsilon)\le 1$. We can equivalently rewrite the particle current in
a form which can be seen as the classical analogue to the Landauer-B\"uttiker approach:
\be
J_\rho=\frac{1}{h}\int_0^\infty d\epsilon\,
[f_L(\epsilon)-f_R(\epsilon)]
{\cal T}(\epsilon),
\ee
where
$f_\alpha(\epsilon)=e^{-\beta_\alpha (\epsilon-\mu_\alpha)}$ is the Maxwell-Boltzmann
distribution function. Similarly, we obtain the heat current from reservoir $\alpha$
as
\be
J_{h,\alpha}=\frac{1}{h}\int_0^\infty d\epsilon\, (\epsilon-\mu_\alpha)
[f_L(\epsilon)-f_R(\epsilon)]
{\cal T}(\epsilon).
\ee

To proceed we take the reference electrochemical potential to be that of reservoir
$L$ and set $\mu_L=0$. Following the same steps as done in Refs.~\cite{whitney1, whitney2}
for the quantum case, we find the transmission function that maximizes the efficiency
of the heat engine, $\eta(P)=P/J_{h,L}$, for a given output power $P=(\Delta \mu) J_\rho$,
with $\Delta\mu=\mu_R-\mu_L>0$ and $P,J_{h,L}>0$. It turns out that the optimal ${\cal T}$
is a boxcar function, ${\cal T} (\epsilon)=1$ for $\epsilon_0<\epsilon<\epsilon_1$ and
${\cal T}(\epsilon)=0$ otherwise. Here $\epsilon_0=\Delta\mu/\eta_C$ is obtained from the
condition $f_L(\epsilon_0)=f_R(\epsilon_0)$ and $\epsilon_1$ can be determined numerically
by solving the equation $\epsilon_1=\Delta\mu J_{h,L}^\prime/P^\prime$, where the prime
indicates the derivative over $\Delta\mu$ for fixed ${\cal T}$ (this equation is
transcendental since $J_{h,L}$ and $P$ depend on $\epsilon_1$). The maximum achievable
power (according to scattering theory) is obtained when $\epsilon_1\to\infty$:
\be
P_{\rm max}^{({\rm st})}=A\frac{\pi^2}{h}\,k_B^2\,(\Delta T)^2,
\label{eq:pendry}
\ee
where $\Delta T=T_L-T_R$ and $A\approx 0.0373$. Note that $\Delta\mu$ is determined
from the above optimization procedure; in particular at $P_{\rm max}^{({\rm st})}$ we
obtain $\Delta\mu=k_B\Delta T$. At small output power, $P/P_{\rm max}^{({\rm st})}\ll 1$,
the upper bound on efficiency approaches the Carnot efficiency as follows:
\be
\eta(P)\le \eta_{\rm max}^{({\rm st})}(P)=\eta_C\left(
1-B\sqrt{\frac{T_R}{T_L}\frac{P}{P_{\rm max}^{({\rm st})}}}
\right),
\label{eq:whitney}
\ee
where $B\approx 0.493$. In the limit $\epsilon_1\to\epsilon_0$, $P\to 0$ and
$\eta\to\eta_C$. In this case, we recover the well-known delta-energy filtering
mechanism to achieve the Carnot efficiency~\cite{mahan, linke1, linke2}. Namely, we
recover the Carnot limit when transmission is possible only inside an energy window of
width $\delta=\epsilon_1-\epsilon_0\to 0$. It is intuitive that selecting transmission
over a tiny energy window greatly reduces power production. It is therefore natural to
expect that a different mechanism to reach Carnot efficiency might allow a larger power
production. Indeed  in what follows we show that for interacting, momentum-conserving
systems, where the Carnot efficiency can be reached without delta-energy filtering
(see Sec. C in the supplementary material), a greatly improved power-efficiency
trade-off can be obtained.

{\it Momentum-conserving systems.}
We consider a system of elastically colliding particles, in contact with two
reservoirs tuned at different temperatures and electrochemical potentials in order
to maintain a steady flow of particles and heat. The equations connecting fluxes and
thermodynamic forces within linear response (an approximation that we will show later
to be valid for our model) are~\cite{callen, degrootmazur}
\begin{equation}
\left(
\begin{array}{c}
J_\rho\\
J_u
\end{array}
\right) = \left(
\begin{array}{cc}
L_{\rho \rho} & L_{\rho u} \\
L_{u \rho} & L_{u u}
\end{array}
\right) \left(
\begin{array}{c}
-\nabla(\beta\mu)\\
\nabla \beta
\end{array}
\right) ,
\label{eq:lresponse}
\end{equation}
where $J_\rho$ is the steady particle current, $J_u$ is the steady energy current,
and $L_{ij}$ (with $i,j=\rho,u$) are the kinetic (Onsager) coefficients. Hereafter
we will discuss our results in the language of thermoelectricity, even though they
could equally well refer to other steady-state heat to work conversion phenomena
like thermodiffusion. The Onsager coefficients are then related to the familiar
transport coefficients as follows:
\begin{equation} \label{transport}
\sigma=\frac{e^2}{T}\,L_{\rho\rho},
\quad\kappa=\frac{1}{T^2}\frac{\det\mathbb{L}}{L_{\rho\rho}},
\quad S=\frac{1}{eT}\left(\frac{L_{\rho u}}{L_{\rho\rho}}-\mu\right).
\end{equation}
Here $\sigma$ is the electrical conductivity, $\kappa$ is the thermal conductivity,
and $S$ is the thermopower; Besides, $e$ is the charge of the conducting particles,
$T\approx T_L\approx T_R$ and $\mu\approx \mu_L\approx \mu_R$ in the linear response
formulas, and $\det\mathbb{L}$ denotes the determinant of the (Onsager) matrix of
kinetic coefficients. Thermodynamics imposes $\det \mathbb{L}\ge 0$,
$L_{\rho\rho}\ge 0$, $L_{uu}\ge 0$, and the Onsager reciprocity relations ensure
(for systems with time-reversal symmetry) that $L_{u\rho} =L_{\rho u}$. The maximum
efficiency for energy conversion achievable by the system is a monotonically growing
function of the thermoelectric figure of merit $ZT$~\cite{Benenti2017}:
\begin{equation} \label{ZT}
ZT=\frac{\sigma S^2}{\kappa}\,T=
\frac{(L_{u\rho}-\mu L_{\rho\rho})^2}{\det \mathbb{L}}.
\end{equation}
Thermodynamics imposes $ZT\ge 0$, with the efficiency $\eta=0$ when $ZT=0$ and
$\eta\to\eta_C$ when $ZT\to\infty$.

\begin{figure}[!t]
\includegraphics[width=9.0cm]{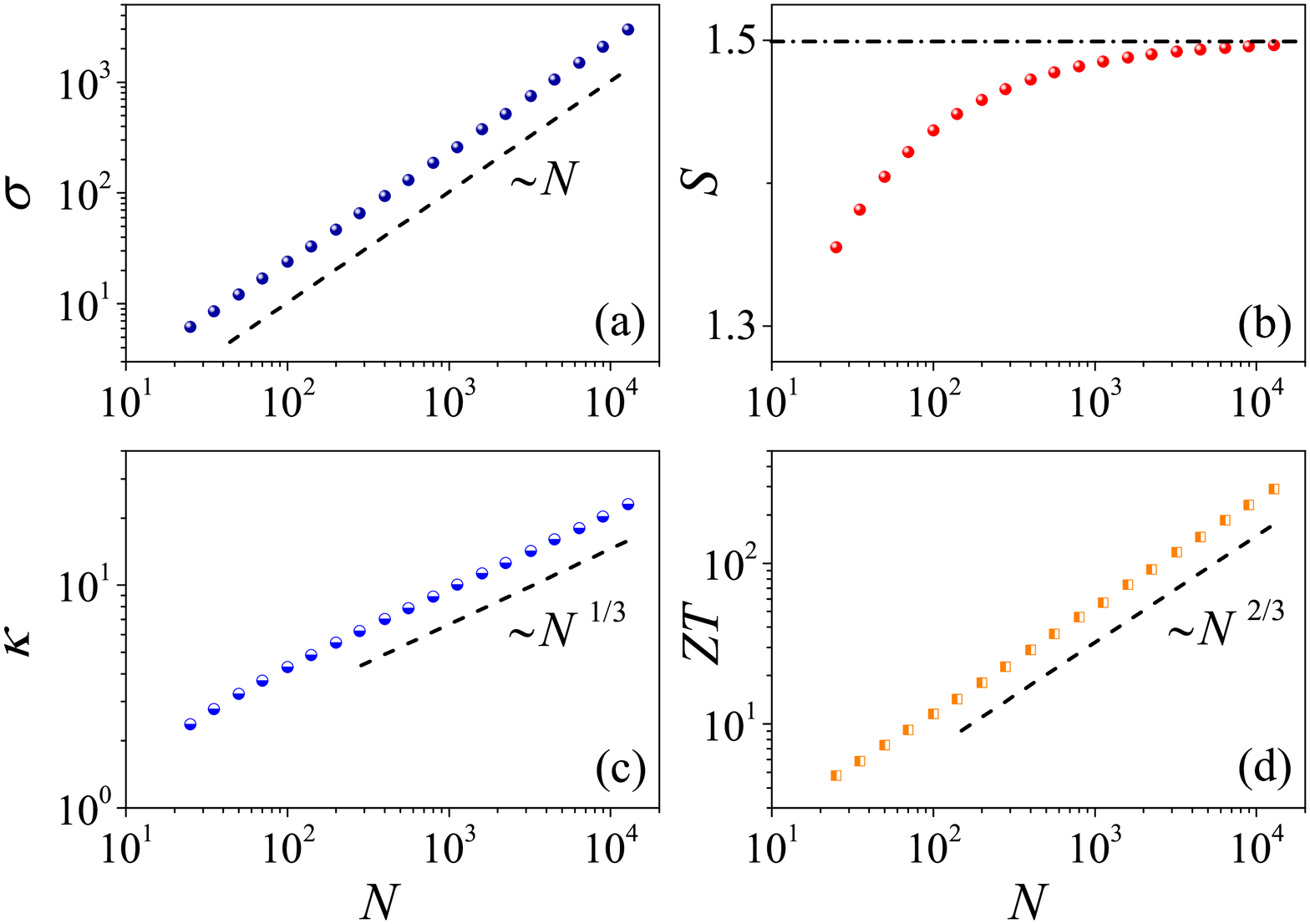}
\caption{Electrical conductivity $\sigma$ (a), thermopower $S$ (b), thermal
conductivity $\kappa$ (c) and figure of merit $ZT$ (d) as a function of the mean
number $N$ of particles inside the system, for the one-dimensional, diatomic
hard-point gas. Here and in the other figures, the data are obtained for $M=3$,
$T=1$, and $\mu=0$.}
\label{fig:ZT}
\end{figure}

Hereafter, we illustrate the breaking of bound (\ref{eq:whitney}) by considering
a one-dimensional, diatomic chain of hard-point elastically colliding particles
connected to reservoirs, with masses $m_i\in\{m,M\}$ and $m\ne M$.
(See~\cite{saito2010} for details of the model.) We have performed a nonequilibrium
calculation of the transport coefficients and then of the figure of merit $ZT$
(we have developed a method to determine very accurately the transport coefficients,
see the supplementary material, Sec.~B,  for details). In our simulations, we set
$k_B=m=e=1$ and the system length, $L$, to be equal to the mean number of particles,
$N$, inside the system. Our data shown in Fig.~\ref{fig:ZT} as well as theoretical
arguments~\cite{Benenti2013} show that the electrical conductivity $\sigma\propto N$,
the thermal conductivity $\kappa\propto N^\xi$, with the power $\xi=1/3$ predicted
by hydrodynamics approach~\cite{lepri,dhar}, the thermopower is asymptotically
size-independent, and therefore $ZT\propto N^{1-\xi}=N^{2/3}$.

\begin{figure}[!t]
\includegraphics[width=8.9cm]{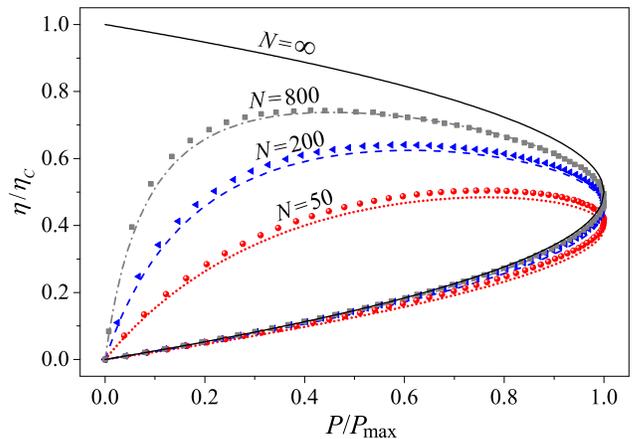}
\caption{Relative efficiency $\eta/\eta_C$ versus normalized power $P/P_{\rm max}$
for $\Delta T=0.2$ ($T_L=1.1$, $T_R=0.9$) and different system sizes. The dotted,
dashed, dot-dashed curves show the expectation from linear response, Eq.~(\ref{eq:loops}),
at the $ZT(N)$ value corresponding to the given system size $N$. The solid line is
Eq.~(\ref{eq:loops}) for $ZT=\infty$, corresponding to $N=\infty$ in our model. The
upper branch of this curve sets the linear-response upper bound on efficiency for
a given power.}
\label{fig:loops}
\end{figure}

In Fig.~\ref{fig:loops}, we show, for a given $\Delta T$ and different system
sizes, the relative efficiency $\eta/\eta_C$ as a function of the normalized power
$P/P_{\rm max}$. Note that these curves have two branches as they are obtained by
changing $\Delta \mu$ from zero (where $P=0$) up to the stopping value, where again
$P=0$, since the electrochemical potential difference becomes too high to be overcome
by the temperature difference. In between, power first increases, up to $P=P_{\rm max}$,
and then decreases, leading to a two-branch curve. In the same figure, we also show the
analytical result from linear response~\cite{Benenti2017}:
\begin{equation}
\frac{\eta}{\eta_{\rm C}}=
\frac{\displaystyle{\frac{P}{P_{\rm max}}}}{
\displaystyle{2\left(1+\frac{2}{ZT}\mp \sqrt{
1-\frac{P}{P_{\rm max}}}\right)}},
\label{eq:loops}
\end{equation}
where the figure of merit $ZT$ and $P_{\rm max}=S^2 \sigma (\Delta T)^2/(4N)$, derived
from Eq.~(\ref{eq:lresponse}) and (\ref{transport}), have been computed previously (see 
Fig.~\ref{fig:ZT}). In spite of the not so small value of $\Delta T/T=0.2$, there is a 
good agreement between the results of our numerical simulations and the universal linear 
response behavior given by Eq.~(\ref{eq:loops}). Moreover, such agreement improves with 
increasing the system size, as expected since $|\nabla T|=\Delta T/N$ decreases when $N$ 
increases. For any given $\Delta T$, we expect the linear response to correctly describe 
the transport properties of our model for large enough system sizes. In
Fig.~\ref{fig:loops}, we also show the parabolic curve corresponding to
Eq.~(\ref{eq:loops}) for $ZT=\infty$ (obtained in our model in the thermodynamic limit
$N\to\infty$), whose upper branch is the universal linear response upper bound to
efficiency for a given power $P$. The expansion of such curve for $P/P_{\rm max}\ll 1$
leads to
\be
\eta(P)\le \eta_{\rm lr}(P)=\eta_C\left(1-\frac{1}{4}\,\frac{P}{P_{\rm max}}
\right),
\label{eq:etalr}
\ee
which sets a much less restrictive bound for efficiency-power trade-off than the
bound (\ref{eq:whitney}) obtained above for non-interacting systems. Our above
reported numerical results strongly suggest that the linear-response bound is
saturated by our model in the \emph{thermodynamic limit}.

\begin{figure}[!t]
\includegraphics[width=8.9cm]{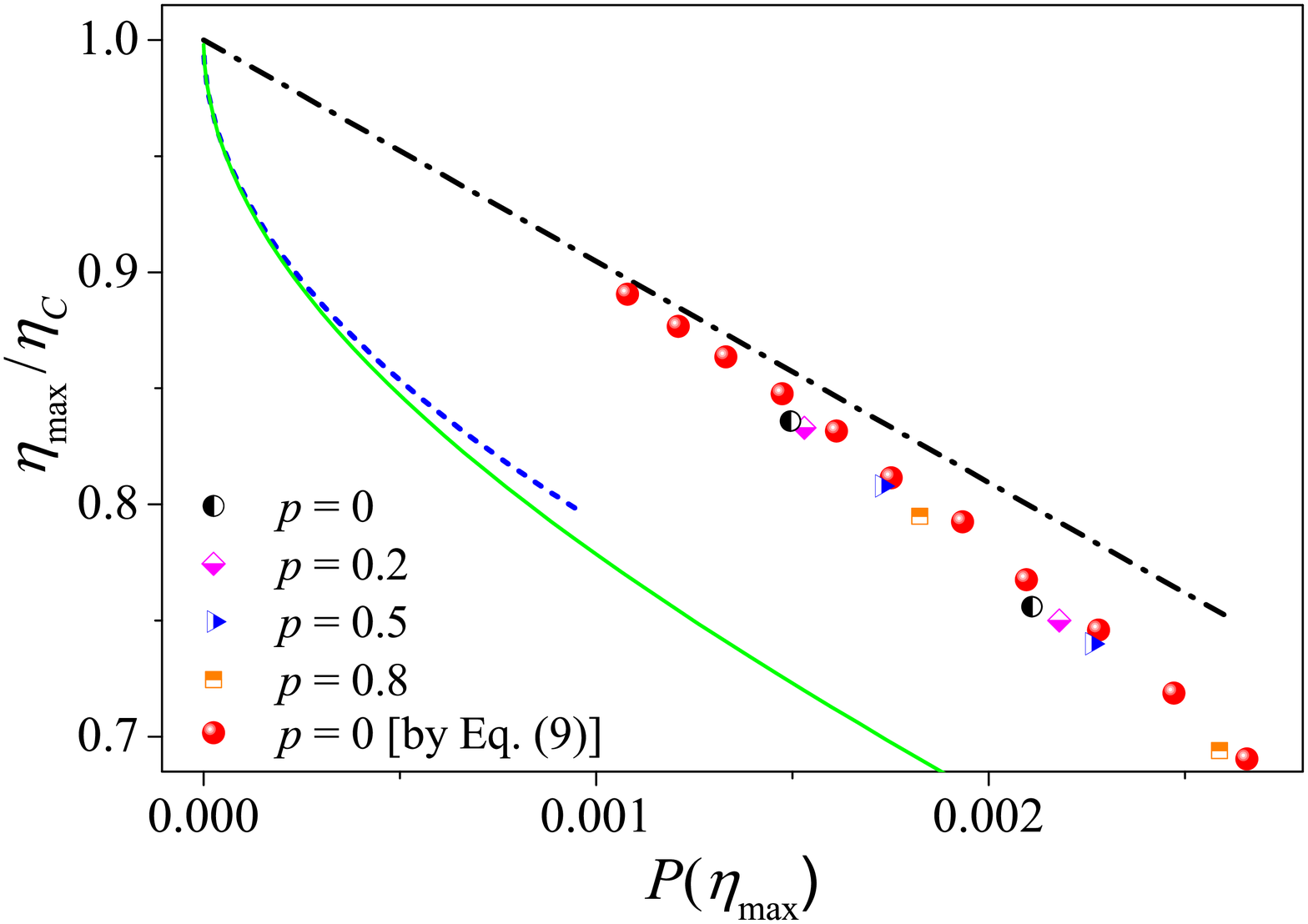}
\caption{Maximum efficiency $\eta_{\rm max}$ versus the corresponding power
$P(\eta_{\rm max})$, from linear response with the values of $ZT$ obtained
numerically (red circles) and directly from numerical computation of power and
efficiency (black and white circles) for various system sizes with $T_L=1.1$ and
$T_R=0.9$. The dot-dashed line is for the analytical expectation from linear
response at large $ZT$, Eq.~(\ref{eq:etaPstar}). We also show the bound for
classical non-interacting systems (solid line) and its approximation at the low
power limit given by Eq.~(\ref{eq:whitney}) (dashed line) as a comparison. Data
from the stochastic model described in the text, with $p$ collision probability
each time two particles meet, are also reported (for further details on this model,
see Sec. A in the supplementary material).}
\label{fig:etapstar}
\end{figure}

To illustrate the breaking of bound (\ref{eq:whitney}) for \emph{finite} system
sizes, we compute the maximum efficiency $\eta_{\rm max}$ and the corresponding
power $P(\eta_{\rm max})$, for different system sizes. The obtained results, shown
as black-white circles in Fig.~\ref{fig:etapstar}, are in agreement with the linear
response predictions, obtained from Eq.~(\ref{eq:loops}) at different values of
$ZT$ (red circles). For $ZT\to\infty$ (obtained when $N\to\infty$), $\eta_{\rm max}
\to\eta_C$ and $P(\eta_{\rm max})\to0$. In the limit of large $ZT$, from
Eq.~(\ref{eq:loops}) we obtain
\be
\eta_{\rm max}=\eta_C\left(1-\frac{1}{2}\,\frac{P(\eta_{\rm max})}{P_{\rm max}}\right).
\label{eq:etaPstar}
\ee
This power-efficiency trade-off when approaching the Carnot efficiency is much
more favorable than the bound for non-interacting systems, also shown for comparison
in Fig.~\ref{fig:etapstar}. To investigate the dependence of power and efficiency on
the interaction strength, we introduce a parameter $p$ as follows: When two particles
meet, they pass through each other with probability $p$, while they collide elastically
with probability $1-p$. For our original hard-point model $p=0$, while for the
noninteracting case $p=1$. We can see in Fig.~\ref{fig:etapstar} that data at different
values of $p$ stay on a single curve, as expected from linear response. While for a
given system size by decreasing interactions (i.e., by increasing $p$) we reduce $ZT$
and therefore deteriorate the performance of energy conversion, $ZT$ still grows with
the system size. In short, the larger $p$ the larger the system size is required to
have a given number of collisions per particle crossing the system. Only in the
non-interacting case we obtain $ZT=1$ ($\eta/\eta_C\approx 0.17$) for all system
sizes~\cite{Benenti2013}.

{\it Conclusions and discussion.}
In this letter, we have shown that classical interacting systems allow, for a given
power, a much higher efficiency than the one achievable in the non-interacting case.
This result shows that interactions can significantly improve the performance of
heat to work conversion. Our results are based on the fact that for momentum-conserving
systems the Carnot efficiency can be achieved at the thermodynamic limit without
delta-energy filtering. While we have considered for illustrative purposes a
one-dimensional, diatomic disordered chain of hard-point elastically colliding
particles, our theoretical considerations can be as well extended to other
momentum-conserving systems, also of higher dimensions~\cite{Benenti2014, Chen2015}.
In the non-interacting case, for $d$-dimensional systems connected to reservoirs via
openings of linear size $l_\alpha$, the injection rate of particles from reservoir
$\alpha$ to the system is proportional to $(l_\alpha/\lambda_\alpha)^{d-1}$, and
therefore the maximum power scales linearly with this quantity, which plays the
role of the number of transverse modes in a classical context. The corresponding
non-interacting bound on efficiency at a given power is broken by momentum-conserving
systems, as shown in the supplementary material, Sec.~D, for the two-dimensional
multi-particle collision model~\cite{kapral}. Finally, we have also considered
refrigeration (see again the supplementary material, Sec.~E) and shown that, thanks to
interactions, one can greatly exceed the bound on efficiency for a given cooling power
which applies to systems described by the scattering theory. While we conjecture that
our results also apply in the quantum case for systems with momentum conservation,
such extension remains as a challenging task for future investigations.

Besides their fundamental interest, our findings for momentum-conserving systems could
be of practical relevance in situations where the elastic mean free path of the conducting
particles is much longer than the length scale over which interactions are effective
in exchanging momenta between the particles, as it might happen in high-mobility
two-dimensional electron gases at low temperatures. Moreover, our results might find
applications in the context of cold atoms, where a thermoelectric heat engine has
already been demonstrated for weakly interacting particles~\cite{brantut}.
More recent experimental results on coupled particle and heat transport through a
quantum point contact connecting two reservoirs of interacting Fermi gases have shown
a strong violation of the Wiedemann-Franz law which could not be explained by the
Landauer-B\"uttiker scattering theory~\cite{esslinger}. It can be envisaged that
in such systems, which can be considered as thermoelectric devices with high
efficiency~\cite{esslinger}, the non-interacting bound on efficiency for a given
(cooling) power could be outperformed, with possible applications to the refrigeration
of atomic gases.


\appendix
\section{Supplementary material}

\subsection{Numerical study of efficiency and power}

\begin{figure*}[!]
\includegraphics[width=18.2cm]{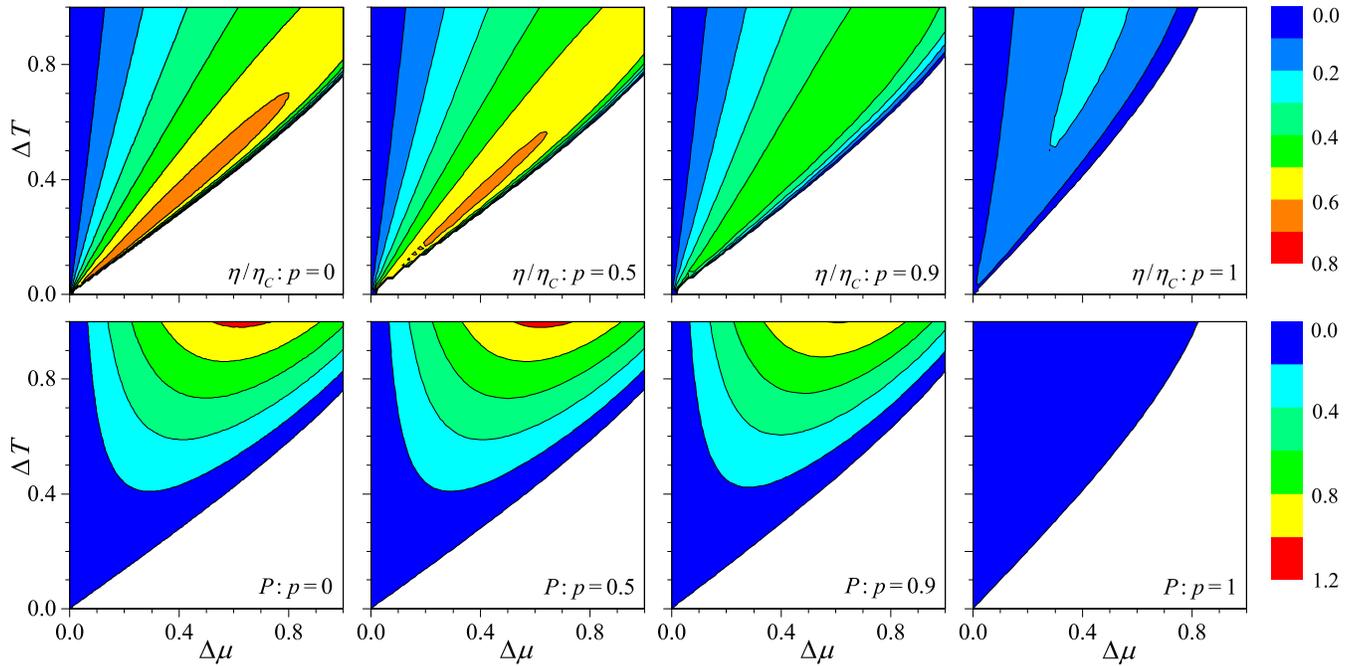}
\caption{The numerically obtained efficiency (normalized to the Carnot
efficiency, upper row) and the power (lower row) as a function of the
temperature difference $\Delta T$ and the electrochemical potential
difference $\Delta \mu$ between two applied reservoirs for the hard-point
diatomic chain model (first column) and its variants parameterized by a
non-zero $p$ (from the second to the forth column) for $T=1$, $\mu=0$,
and $L=N=200$. The parameter $p$ introduces a stochastic element in the
systems's dynamics: when two particles meet, they may (with a probability
$p$) pass through each other instead of colliding. All four panels in each
row are plotted with the same scale indicated at the end. In the white area
of each panel the power is negative, and therefore the system cannot be used
for power production.}
\label{fig:eta}
\end{figure*}

The efficiency and the power of the illustrating one-dimensional (1D) diatomic
hard-point chain model have been thoroughly investigated by molecular dynamics
simulations. The setup consists of two reservoirs at temperature $T_L=T+\Delta
T/2$ and $T_R=T-\Delta T/2$ and electrochemical potential $\mu_L=\mu-\Delta
\mu/2$ and $\mu_R=\mu+\Delta \mu/2$ (with $\Delta T>0$ and $\Delta \mu>0$),
respectively, applied to the system at the two ends. The system length, $L$,
is set to be equal to the averaged particle number, $N$, in the system;
therefore in the following we refer to the system size by $L$ or $N$
equivalently. Initially, the system is evolved till it relaxes to the stationary
state, then the time averaged particle current $J_\rho$ and energy current $J_u$
are evaluated by evolving the system further. Finally, the power and efficiency
are obtained by definition: $P=\Delta \mu J_\rho$ and $\eta=P/J_{h,L}$, where
$J_{h,L}=J_u-\mu_L J_\rho$ is the heat current flows from the left (hotter)
reservoir. As an illustration, the first column in Fig.~\ref{fig:eta} shows
the typical results.

Note that our model is interacting and nonintegrable. If particles do not
interact, i.e., when two particles meet they simply pass through each other
without collision, then the system becomes integrable. In order to reveal
how the efficiency and the power depend on the interaction strength -- which
is our main motivation for this work -- we introduce a probability, $p$,
that controls the latter and study the dependence of $\eta$ and $P$ on
it~\cite{acknowledge}. This parameter controls the interaction strength in
such a way: when the two particles meet, they have the chance $p$ to pass
through each other and the chance $1-p$ to collide elastically. For our
original hard-point model $p=0$, while for the noninteracting, integrable
case $p=1$. For intermediate values of $p$ ($0<p<1$) the dynamics is stochastic.
As Fig.~\ref{fig:eta} shows, when the interaction increases ($p$ decreases),
both the efficiency and the power increase. In particular, in the range of
weak interaction the increasing is remarkably fast. When $p$ decreases to
about 0.5, the increasing almost stops and the values of $\eta$ and $P$ are
close to their saturated values at $p=0$. These results clearly show the
important role interactions play in thermoelectric performance.

We have also studied thoroughly the dependence of efficiency and power
on the system size $N$. We find that for $p<1$, for given $\Delta T$ and
$\Delta \mu$, both $\eta$ and $P$ increase with $N$, which is consistent
with the fact that the figure of merit $ZT$ increases with $N$ and agrees
with our theoretical predictions for an interacting system (see Fig.~1
and Fig.~2 in the paper for the case $p=0$). But for the noninteracting,
integrable case $p=1$, as both $J_\rho$ and $J_u$ are independent of the
system size, $\eta$ and $P$ do not depend on $N$ either. In this case $ZT$
can be shown to be an $N$-independent constant ($ZT=1$)~\cite{Benenti2013}.

\subsection{Numerical method for evaluating \\the figure of merit $ZT$}

Numerically, it is a challenge to evaluate the figure of merit $ZT$ in a
momentum conserving system when the system size is large. This is also the
challenge we encounter for the present study. To overcome this difficulty,
we have developed a general, efficient numerical method that we will outline
here. We start from the linear response equations
\begin{equation}
\left(
\begin{array}{c}
J_\rho\\
J_u
\end{array}
\right) = \left(
\begin{array}{cc}
L_{\rho \rho} & L_{\rho u} \\
L_{u \rho} & L_{u u}
\end{array}
\right) \left(
\begin{array}{c}
-\nabla \alpha\\
\nabla \beta
\end{array}
\right) ,
\end{equation}
where $\alpha\equiv \beta\mu$. In the literature, the standard method to determine
the Onsager coefficients~\cite{carlos2001, carlos2003, Benenti2013} is as follows:
First, we set $\Delta \alpha=\Delta$ and $\Delta\beta=0$, then we can obtain
$L_{\rho \rho}=-J_\rho N/\Delta$ and $L_{u \rho}=-J_u N/\Delta$ by calculating the
currents $J_\rho$ and $J_u$ in the simulation. After that, we set $\Delta \alpha=0$
and $\Delta\beta=\Delta$ instead, then we can get $L_{\rho u}=J_\rho N/\Delta$ and
$L_{uu}=J_u N/\Delta$ in the same way. Based on these Onsager coefficients, we can
derive the transport coefficients $\sigma$, $\kappa$, and $S$, and finally the
thermoelectric figure of merit $ZT$. This method (hereafter referred to as method
A) requires that $\Delta$ is small enough to obtain results independent of $\Delta$,
as expected for the linear response. We find that $\sigma$ and $S$ do not depend
on $\Delta$ sensitively, while the problem arises in computation of thermal
conductivity
\be
\kappa=\frac{1}{T^2}\left(
L_{uu}-\frac{L_{\rho u}L_{u \rho}}{L_{\rho\rho}}
\right),
\label{eq:heat}
\ee
as in this case we need to compute a quantity which scales as $\sim N^{1/3}$ from
the difference of two quantities, $L_{uu}$ and $L_{\rho u}L_{u \rho}/L_{\rho\rho}$,
which scale linearly with the system size due to the momentum is conserved~\cite{
Benenti2013}. For a given $\Delta$, the error in such quantities also scales
linearly with $N$, and therefore with method A it is possible to measure $\kappa$
accurately only if we take $\Delta \sim 1/N$. This implies prohibitive costs in
the simulations for large $N$.

\begin{figure}
\includegraphics[width=8.9cm]{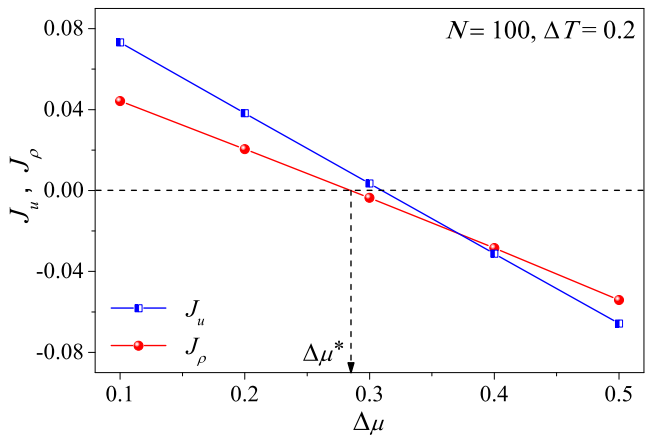}
\caption{Particle and energy current as a function of $\Delta \mu$ for a given
$\Delta T$ in the hard-point diatomic chain model.}
\label{fig:methods1}
\end{figure}

To solve this problem, we have developed a different method (refereed to as method
B) that allows us to compute $\kappa$ accurately for a fixed $\Delta$ at any value
of $N$. Since the heat conductivity is computed at zero particle flow, the key point
is to find values $(\Delta \alpha, \Delta \beta)$, or equivalently $(\Delta T,
\Delta \mu)$, which ensure $J_\rho=0$. Our method is as follows:
\begin{itemize}
\item
For a given value of $\Delta T$, set $T_L=T+\Delta T/2$ and $T_R=T-\Delta T/2$.
Then calculate $J_\rho$ and $J_u$ by simulations as a function of $\Delta\mu$
with $\mu_L =\mu-\Delta\mu/2$ and $\mu_R=\mu+\Delta\mu/2$;
\item
Based on the function $J_\rho(\Delta\mu)$, determine $\Delta\mu^\ast$ such that
$J_\rho(\Delta\mu^\ast)=0$ , then the heat conductivity is evaluated from the
relation
\be
J_u(\Delta\mu^\ast)=-\kappa \nabla T= \kappa \Delta T/N.
\ee
\end{itemize}
Figure~\ref{fig:methods1} shows the typical dependence of $J_\rho$ and $J_u$ on
$\Delta\mu$. Interpolating the data points for $J_\rho$ we can determine $\Delta
\mu^\ast$ from the condition $J_\rho(\Delta \mu^\ast)=0$ and then evaluate $J_u
(\Delta \mu^\ast)$ and $\kappa$ as explained above. Note that the interpolation
procedure can be iterated to obtain more accurate value of $\Delta\mu^\ast$ and
consequently of $\kappa$.

\begin{figure}
\includegraphics[width=8.9cm]{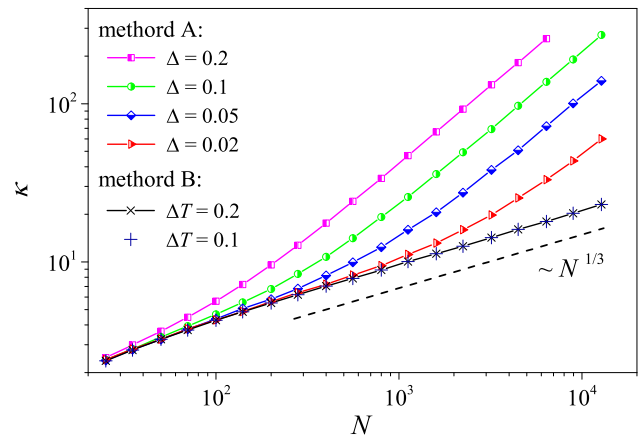}
\caption{Comparison of the standard method used in the literature for computing
the heat conductivity (method A) with that we have developed
(methods B).}
\label{fig:methods2}
\end{figure}

We finally compare in Fig.~\ref{fig:methods2} method B for two given values of
$\Delta T$ with method A at different values of $\Delta$. The results of method
A coincide with those of method B, provided $\Delta$ is taken smaller and smaller
when $N$ increases. This shows that method B is reliable and suitable to obtain
accurate results for the thermal conductivity at large system sizes
(in Fig.~\ref{fig:methods2} up to $N=12800$).

\subsection{Energy distribution}

For a thermoelectric system that can be modelled by the scattering theory, the
Carnot efficiency can be achieved for delta-energy filtering, i.e., when only
carriers within a given energy window are allowed to transmit through the system,
and the width of such window tends to zero~\cite{mahan, linke1, linke2}. It is
therefore interesting to investigate if this mechanism also works in our diatomic
chain model~\cite{acknowledge}: Does the energy distribution in the system shrink
as the system size increases?

\begin{figure}[!t]
\includegraphics[width=8.8cm]{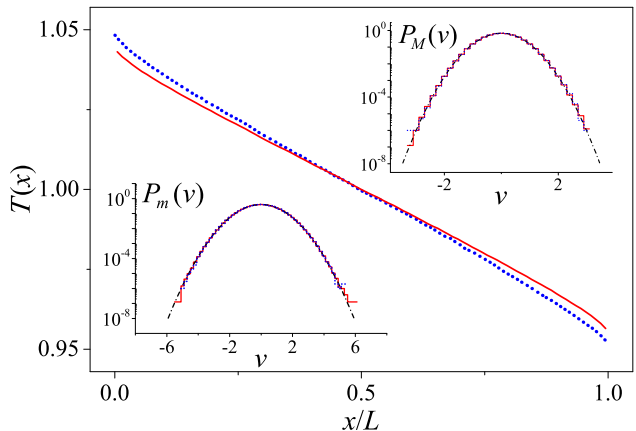}
\caption{The temperature profile of the steady state (main panel) and the velocity
distributions of the two species of particles at the middle point $x=L/2$ (insets)
of the diatomic chain system with $m=1$ and $M=3$. In the main panel and the insets,
the red solid (blue dotted) line is for $L=N=100$ (800). The black dash-dotted line
in the insets are for the Boltzmann distribution for mass $m$ and $M$, respectively,
with the evaluated temperature $T(x)=1$. Note that the three lines in each inset
agree with each other so well that it is difficult to distinguish them.}
\label{fig:vdis}
\end{figure}

To this end, we study the velocity distributions of the two species of particles,
$P_m(v)$ and $P_M(v)$, when they pass a given position, $x$, in the system. We
find that they always agree perfectly with the Boltzmann distribution at a local
temperature $T(x)$ for $m\ne M$. Indeed, due to interactions, our system always
relaxes to a steady state and reaches the local equilibrium, and this property does
not depend on the system size. Therefore the mechanism for the Carnot efficiency
is approached in our system is fundamentally different from the delta-energy filtering.

As an example, in Fig.~\ref{fig:vdis} the velocity distributions and the temperature
profiles at two different system sizes, $L=N=100$ and $800$, are presented and
compared. The temperatures and the electrochemical potentials of the reservoirs
are $T_L=1.05$, $T_R=0.95$, $\mu_L=-\Delta\mu^\ast/2$, and $\mu_R=\Delta\mu^\ast/2$,
respectively, with $\Delta\mu^\ast=0.143$ (0.148) for $N=100$ (800). The system
is evolved till it relaxes to the steady state, then the velocity distributions
along the system are calculated. The simulation results suggest convincingly that
$\langle \frac{1}{2}mv^2\rangle=\langle \frac{1}{2}Mv^2\rangle$ at any position $x$,
which implies the local equilibrium. The temperature can therefore be defined and
obtained by identifying the averaged energy to $\frac{1}{2}k_BT(x)$.

The insets of Fig.~\ref{fig:vdis} show the velocity distributions at the middle
point $x=L/2$ of the system, and they all fully agree with the Boltzmann distribution
of temperature $T(x)=1$. At other places the velocity distributions are also the
same as the Boltzmann distribution with a certain temperature, which is evaluated
and presented in the main panel. Note that the temperatures at the left and right
ends of the system are slightly different from the external temperatures $T_L$ and
$T_R$. This boundary effect, which is the result of a boundary (Kapitza) resistance,
vanishes as the system size increases.

\subsection{An additional illustrating example: The 2D momentum-conserving gas of interacting particles}

\begin{figure}[!t]
\vskip0.45cm
\includegraphics[width=7.1cm]{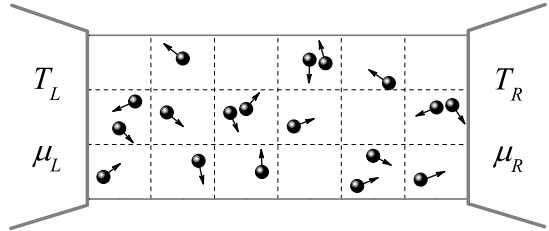}
\vskip0.4cm
\caption{Schematic plot of the 2D momentum-conserving gas of interacting particles,
described by the multi-particle collision dynamics. The cells of dashed-line boundaries
represent the partition of space considered for modeling collisions.}
\label{fig:2DMPC}
\end{figure}

For non-integrable momentum-conserving systems, our theory is general. To show its
generality and in particular its independence of dimensionality, here we provide
another illustrating example. The model system we consider is a two-dimensional
(2D) momentum-conserving gas of interacting particles (see Fig.~\ref{fig:2DMPC} for
a schematic plot). All particles have the same mass $m$, and we set $m=1$. The length
of the system is $L$; two reservoirs are coupled at the two ends. The width of the
system, i.e., its size in the transverse direction, is $l$, and in this direction
the periodic boundary condition is applied.

The evolution of the system is described by the multi-particle collision (MPC)
dynamics~\cite{kapral}, introduced as a stochastic model to study solvent dynamics.
Collisions are modelled by coarse graining the time and space at which interactions
occur, and hence simplifies the numerical simulations of interacting particles.
By MPC, the system evolves in discrete time steps, consisting of free propagation
during a time $\tau$ followed by instantaneous collision events. During the free
propagation period, a particle keeps its velocity ${\mathbf v}_i$ unchanged but
changes its position from ${\mathbf r}_i$ to ${\mathbf r}_i+\tau {\mathbf v}_i$.
For the collisions, the system is partitioned into identical square cells of side
$a$ (see Fig.~\ref{fig:2DMPC}), then the velocities of all particles found in the
same cell are rotated with respect to their center of mass velocity ${\mathbf V}_{CM}$
by two angles, $\alpha$ or $-\alpha$, randomly chosen with equal probability. The
velocity of a particle in the cell is thus updated from ${\mathbf v}_i$ to
${\mathbf V}_{CM}+\hat R^{\pm \alpha}({\mathbf v}_i-{\mathbf V}_{CM})$, where
$\hat R^{\pm \theta}$ is the 2D rotation operator of angle $\theta$.

Note that the MPC dynamics keeps the total momentum and energy conserved. By using
our new numerical method (see Sec.~B), we checked that in this model $ZT$ diverges
as the system size increases~\cite{benenti2014}. As our linear response analysis
is independent of the system's dimension, it is expected that the thermoelectric
power-efficiency trade-off follows the theoretical prediction Eq.~(9) and the
asymptotic relation Eq.~(11) given in the main paper.

\begin{figure}
\includegraphics[width=8.9cm]{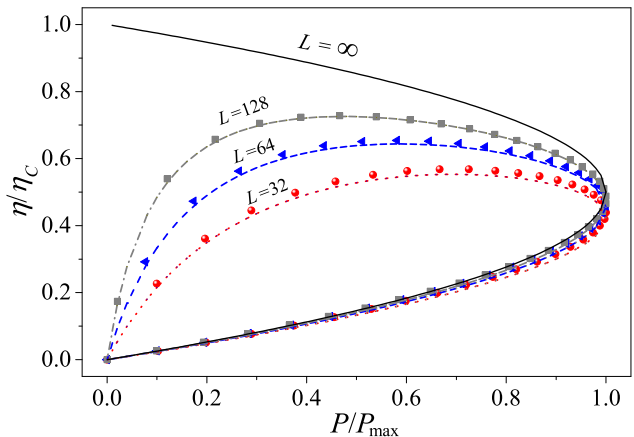}
\caption{Relative efficiency $\eta/\eta_{C}$ versus normalized power $P/P_{\rm max}$
for various system length $L$ of the 2D gas model. The dotted, dashed, dot-dashed
curves show the expectation from the linear response analysis, Eq.~(9) in the main
paper, at the $ZT(L)$ value corresponding to the system length $L$. The solid line
is Eq.~(9) for $ZT=\infty$, corresponding to $L=\infty$. The upper branch of this
curve sets the linear response upper bound on efficiency for a given power.}
\label{fig:2DetaP}
\end{figure}

To verify this conjecture, thorough numerical simulations are carried out for various
values of the system length $L$. Other parameters adopted are as follows: $T_L=1.1$,
$T_R=0.9$, $l=2$, $a=0.1$, $\alpha=\pi/2$, and the averaged particle number density
$\rho=22.75$. The reference electrochemical potential is set to be $\mu=0$ for the
adopted value of $\rho$ at temperature $T=1$. For each given $L$ value, we take
$\mu_L=-\Delta \mu/2$ and $\mu_R=\Delta \mu/2$ and investigate how the thermoelectric
power and efficiency depend on $\Delta \mu$. The results are presented in
Fig.~\ref{fig:2DetaP}; it can be seen that the simulation results agree with the
linear-response theoretical prediction very well.

\begin{figure}
\includegraphics[width=8.9cm]{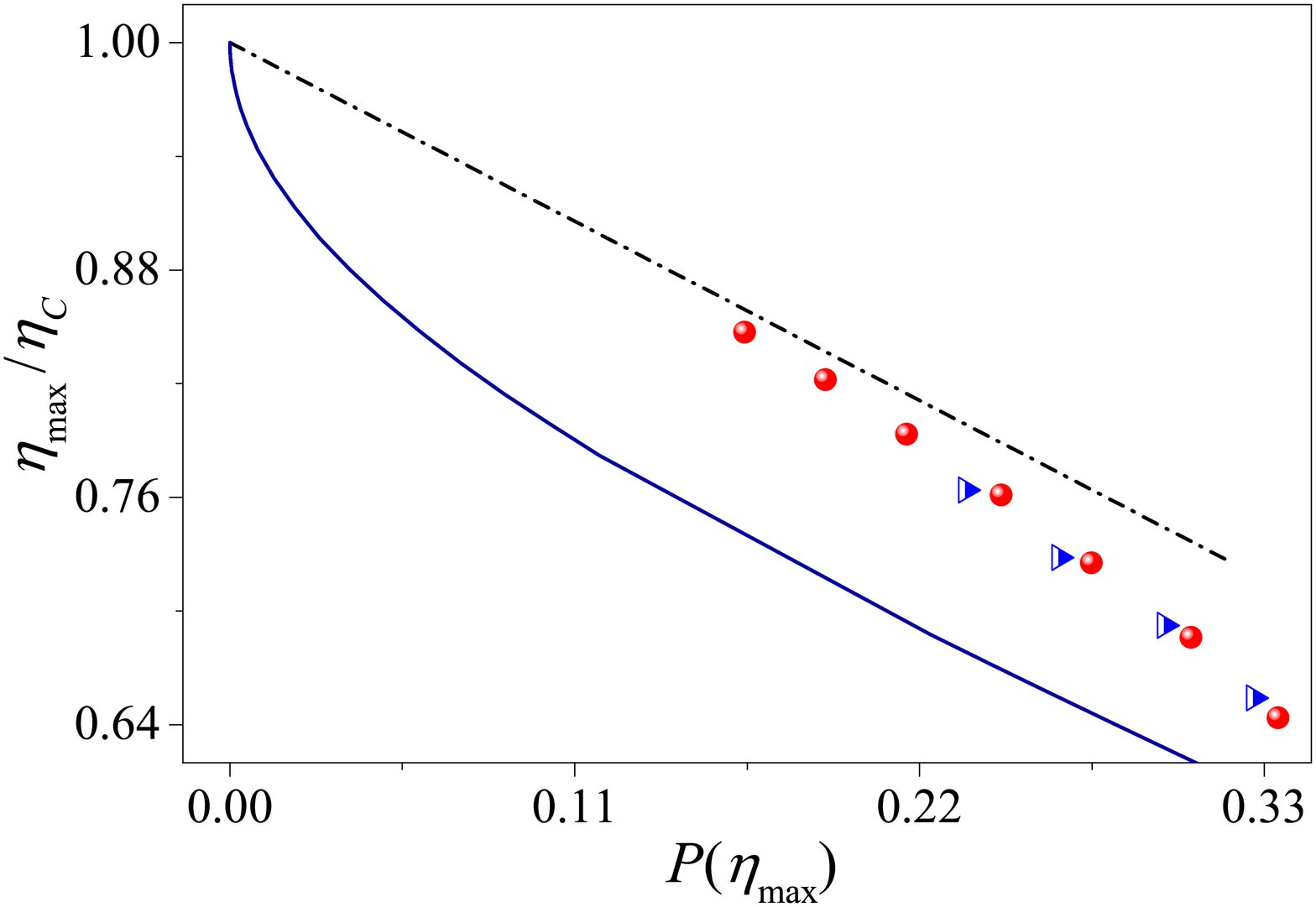}
\caption{Maximum efficiency $\eta_{\rm max}$ versus the corresponding power
$P(\eta_{\rm max})$ of the 2D gas model, evaluated from the linear response theory
[Eq.~(9) in the paper] with the simulated Onsager coefficients (red dots) and
directly from numerical computation of power and efficiency (blue triangles) for
various system sizes. The dot-dashed line is for the analytical expectation from
linear response at large $ZT$, Eq.~(11) in the paper. For comparison, the bound
for 2D classical non-interacting systems (solid line) with $T_L=1.1$, $T_R=0.9$,
and $\rho=22.75$ is also given.}
\label{fig:2Detamax}
\end{figure}

In Fig.~\ref{fig:2Detamax}, we summarize the results for the maximum efficiency
versus the corresponding power for various $L$, obtained via direct simulations
(blue triangles) and via the theoretical result [Eq.~(9) in the paper] with $ZT$
and $P_{\rm max}$ computed from the numerically simulated Onsager coefficients
(red dots). It can be seen that results obtained with the two methods agree with
each other and consistently approach the theoretical prediction [Eq.~(11) in the
paper] for large $ZT$. Importantly, in this figure we can see that the thermoelectric
efficiency at a given power of this interacting system outperforms the 2D
non-interacting bound, which we are going to derive below.

For the 2D non-interacting systems, the maximum efficiency for a given power can
be obtained following the same steps as in the 1D case. The particle reservoirs
are modelled as 2D ideal gas systems. The velocity distribution inside the ideal
gas reservoirs is the Maxwell distribution. Particles enter from reservoir
$\alpha$ into the system through an opening of length $l$, with an injection rate
$\gamma_\alpha$ (for a short description of this effusion process, see Appendix B
in~\cite{Benenti2017}). The particle current is given by
\begin{equation}
J_\rho=
\gamma_L\int_0^\infty d\epsilon u_L(\epsilon) {\cal T} (\epsilon)
-\gamma_R\int_0^\infty d\epsilon u_R(\epsilon) {\cal T} (\epsilon),
\end{equation}
where ${\cal T}(\epsilon)$ is the transmission probability for a particle with
energy $\epsilon$ to transit from one end to the other end of the system,
$0\le {\cal T} (\epsilon)\le 1$. The injection rate from reservoir $\alpha$ is
given by
\begin{equation}
\gamma_\alpha=\frac{l\rho_\alpha}{\sqrt{2\pi m \beta_\alpha}},
\end{equation}
and the energy distribution of the injected particles is
\begin{equation}
u_\alpha(\epsilon)=2\beta_\alpha e^{-\epsilon \beta_\alpha}
\sqrt{\frac{\epsilon\beta_\alpha}{\pi}}.
\end{equation}
We can then express the density and the injection rate in terms of the
electrochemical potential and of the de Broglie thermal wave length
$\lambda_\alpha=h/\sqrt{2\pi m k_B T_\alpha}$ ($m$ is the mass of the
injected particles and $h$ the Planck constant) as
\begin{equation}
\mu_\alpha=k_B T_\alpha \ln(\lambda_\alpha^d \rho_\alpha),
\end{equation}
where $d$ is the dimensionality of the reservoirs and of the system
(in our case, $d=2$). Therefore
\begin{equation}
\gamma_\alpha=\frac{e^{\beta_\alpha\mu_\alpha}}{h\beta_\alpha}
\left(\frac{l}{\lambda_\alpha}\right)^{d-1}.
\end{equation}
We finally obtain that
\be
J_\rho=\frac{2\sqrt{2m}l}{h^2}\int_0^\infty d\epsilon\,\sqrt{\epsilon}\,
[f_L(\epsilon)-f_R(\epsilon)]
{\cal T}(\epsilon)
\label{eq:Jrho2D}
\ee
with $f_\alpha(\epsilon)=e^{-\beta_\alpha(\epsilon-\mu_\alpha)}$. The heat
current from reservoir $\alpha$ is similarly obtained as
\be
J_{h,\alpha}=\frac{2\sqrt{2m}l}{h^2}\int_0^\infty d\epsilon\,\sqrt{\epsilon}\,
(\epsilon-\mu_\alpha)
[f_L(\epsilon)-f_R(\epsilon)]
{\cal T}(\epsilon).
\label{eq:Jh2D}
\ee
Starting from Eqs.~(\ref{eq:Jrho2D}) and (\ref{eq:Jh2D}) for particle and heat
currents in two dimensions, one can follow the 1D derivation reported in the
main text and obtain the 2D bound from scattering theory for the maximum
efficiency at a given power output (shown as the full blue curve in
Fig.~\ref{fig:2Detamax}).

We note that the same derivation can be performed also for three-dimensional
systems, with in that case, for an opening of area $l^2$,
\begin{equation}
u_\alpha(\epsilon)=\beta_\alpha^2 e^{-\epsilon \beta_\alpha}\epsilon,
\end{equation}
and
\be
\gamma_\alpha u_\alpha(\epsilon)=\frac{2\pi m l^2}{h^3}\epsilon
f_\alpha(\epsilon).
\ee

\subsection{Overcoming the scattering-theory bound for refrigeration}

When a thermoelectric device works as a refrigerator, the most important
benchmark is the coefficient of performance (COP)
\be
\eta^{(r)}=\frac{J_{h,L}}{P_{\rm abs}},
\ee
given by the ratio of the cooling power $J_{h,L}$, that is, the heat current
extracted from the cold reservoir (which without lack of generality we assume
to be the left one, i.e., $T_L<T_R$), over the absorbed power $P_{\rm abs}$.
The COP can never exceed Carnot's limit,
\be
\eta^{(r)}\le \eta^{(r)}_C=\left(\frac{T_R}{T_L}-1\right)^{-1}.
\ee
With a calculation analogous to the one performed in the main paper for power
production, we can compute the maximum COP allowed by nonlinear classical
scattering theory for a given cooling power. For 1D systems, we obtain that
the optimal transmission function is a boxcar function,
${\cal T}
(\epsilon)=1$ for $\epsilon_0<\epsilon<\epsilon_1$ and ${\cal T}(\epsilon)=0$
otherwise. Differently from power production, here
$\epsilon_1=-\Delta\mu \eta^{(r)}_C$ ($\Delta\mu=\mu_R-\mu_L<0$) and
$\epsilon_0=-\Delta\mu J_{h,L}^\prime/P_{\rm abs}^\prime$,
where the prime indicates the derivative over $\Delta\mu$
for fixed ${\cal T}$. The maximum cooling power is obtained when
$\epsilon_1\to\infty$ and is given by
\be
(J_{h,L})_{\rm max}^{({\rm st})}=\frac{k_B^2 T_L^2}{h}.
\ee
Note that in this case $-\Delta \mu\to \infty$ as well and therefore the
absorbed power $P_{\rm abs}=-(\Delta\mu) J_\rho\to \infty$, implying that
the COP vanishes. At low cooling power,
$J_{h,L}\ll (J_{h,L})_{\rm max}^{({\rm st})}$, the upper bound on the COP
approaches the Carnot's limit as follows:
\be
\eta^{(r)}\le \eta_{\rm max}^{(r,{\rm st})}(J_{h,L})=
\eta_C^{(r)}\left(
1-C\sqrt{\frac{T_R}{T_R-T_L}\frac{J_{h,L}}{(J_{h,L})_{\rm max}^{({\rm st})}}}
\right),
\label{eq:scattering}
\ee
with $C\approx 0.813$. The Carnot's limit is obtained for delta-energy filtering,
$\epsilon_1-\epsilon_0\to 0$, and in this limit the cooling power vanishes.

Interestingly and importantly, we find that the cooling performance of an
interacting system can surpass the bound set by the nonlinear classical
scattering theory as well at the large $ZT$ regime. By a linear response
analysis~\cite{Benenti2017}, the maximum COP is
\be
\eta_{\rm max}^{(r)}=\eta_C^{(r)}\frac{\sqrt{ZT+1}-1}{\sqrt{ZT+1}+1},
\label{eq:56A}
\ee
which is reached at the cooling power
\be
J_{h,L}=\frac{\Delta T}{L}\sqrt{\kappa(\kappa+S^2\sigma T)},
\label{eq:56B}
\ee
where $\Delta T=T_R-T_L>0$ and $L$ is the system length. At the low-power limit,
it is approximated by
\be
\eta_{\rm max}^{(r)}\approx \eta_C^{(r)}\left(1-\frac{2L J_{h,L}}
{S^2\sigma T \Delta T}\right).
\label{eq:56C}
\ee
As here $\eta_{\rm max}^{(r)}$ is a linear function of $J_{h,L}$, this bound
is, in the high-efficiency region, higher than that of the nonlinear classical
scattering theory, Eq.~(\ref{eq:scattering}).

\begin{figure}
\includegraphics[width=8.9cm]{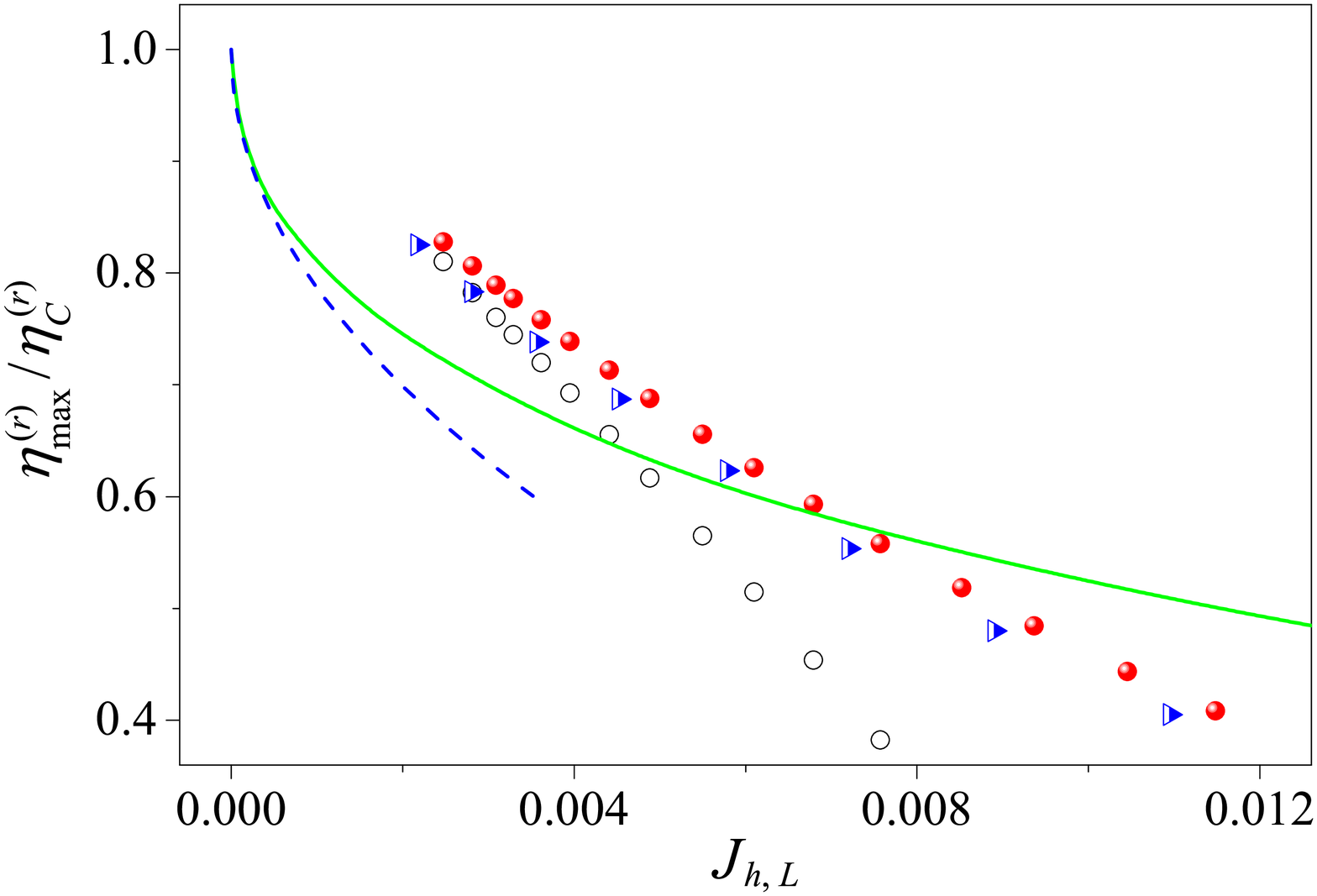}
\caption{The cooling performance of the 1D hard-core, diatomic gas model for
various system sizes with $T_L=0.975$ and $T_R=1.025$. The full red dots are
for the linear response prediction [Eq.~(\ref{eq:56A})-(\ref{eq:56B})] and the
open circles are for its low-power approximation [Eq.~(\ref{eq:56C})]. The blue
triangles are for the direct simulation results. For these three sets of data
points, the system size and $ZT$ increases from right to left. The upper bound
predicted by the nonlinear classical scattering theory and its low-power
approximation [Eq.~(\ref{eq:scattering})] are shown by the solid green line
and the blue dashed line, respectively.}
\label{fig:cooling}
\end{figure}

In order to make a comparison of the two theoretical predictions, we take the 1D diatomic,
hard-core interacting gas again as an illustrating example. We assume that two particles will
not pass through each other when they collide (i.e., $p=0$) and the electrochemical potential at the
studied state of $T=1$ and particle density $\rho=N/L=1$ is zero. The masses of particles are
$m=1$ and $M=3$. First, by using the new numerical method described in Sec.~B, we calculate
the transport coefficients and $ZT$ at the studied state for various system sizes. Then the
dependence of the COP on the cooling power given by the linear response analysis
[Eq.~(\ref{eq:56A})-(\ref{eq:56B})] and its low-power approximation [Eq.~(\ref{eq:56C})]
are evaluated. The results for $T_L=0.975$ and $T_R=1.025$ are presented in
Fig.~\ref{fig:cooling}. It can be seen that the maximum efficiency of the system for a given
cooling power becomes higher than the bound of Eq.~(\ref{eq:scattering}) for noninteracting
systems when $J_{h,L}<0.006$.

In Fig.~(\ref{fig:cooling}), the cooling performance measured in direct simulations for
various system sizes is also presented and the results corroborate our linear response analysis
convincingly. In our simulations for a given system size, the system is coupled to two reservoirs
at temperature $T_L$ and $T_R$ and at electrochemical potential $\mu_L=0$ and $\mu_R=\Delta
\mu$, respectively. For a given value of $\Delta \mu$, the cooling power and the efficiency are
measured in the stationary state. Then by changing $\Delta \mu$, the cooling power and the
efficiency as a function of $\Delta \mu$ are obtained, based on which the maximum efficiency
and the corresponding cooling power are in turn identified.

Our analysis of refrigeration can be extended to two and three dimension straightforwardly.
For the 2D interacting gas with the MPC dynamics (see Sec.~D), we have obtained qualitatively
the same results.

{\it Acknowledgments:} We are grateful to Dario Poletti for fruitful discussions
and to an anonymous referee for useful suggestions. We acknowledge support by NSFC
(Grants No. 11535011 and No. 11335006) and by the INFN through the project QUANTUM.


\end{document}